\documentstyle[12pt]{article}
\input math_macros

\def\etal{\hbox{et al.}}

\def\blank#1#2#3{                     #1 (19#3) #2}
\def\genjour#1#2#3#4{       #1\ \blank{#2}{#3}{#4}}

\def\cpc#1#2#3{   \genjour {Comp.\ Phys.\ Comm.}            {#1} {#2} {#3}}

\def\np#1#2#3{    \genjour {Nucl.\ Phys.}                   {#1} {#2} {#3}}
\def\pl#1#2#3{    \genjour {Phys.\ Lett.}                   {#1} {#2} {#3}}
\def\pr#1#2#3{    \genjour {Phys.\ Rev.}                    {#1} {#2} {#3}}

\def\prl#1#2#3{   \genjour {Phys.\ Rev.\ Lett.}             {#1} {#2} {#3}}

\def \bar{\overline}

\def \arrow{\rightarrow}

\def \tablecaption#1#2{\begin{center}
                       \begin{tabular}{p{1in}p{4in}}
                       Table #1:  & #2
                       \end{tabular}
                       \end{center}
                       \vskip .2in}

\newfont{\bff}{cmtcsc10 scaled \magstep 1}

\def \tablecaption#1#2{\begin{center}
                       \begin{tabular}{p{1in}p{4.5in}}
                       Table #1:  & #2
                       \end{tabular}
                       \end{center}
                       \vskip .2in}

\def \explanatory#1#2 {\parbox{1.5in}{#1} \hfill \parbox{4.5in}{#2}}  

\begin{document}
\begin{titlepage}
\begin{center}
May 25, 1995     \hfill    \begin{tabular}{l}
                             LBL-37224 \\
                             UCB-PTH-95/14 \\
                         \end{tabular}

\vskip .5in

{\large \bf $D$-Meson Mixing in Broken $SU(3)$}\footnote{This work was
supported by the Director, Office of Energy Research, Office of High
Energy and Nuclear Physics, Division of High Energy Physics of the
U.S. Department of Energy under Contract DE-AC03-76SF00098.}
\vskip .50in

\vskip .5in
Thomas A.~Kaeding\footnote{Electronic address:  {\tt kaedin@theorm.lbl.gov}.}

{\em Theoretical Physics Group\\
     Lawrence Berkeley Laboratory\\
     University of California\\
     Berkeley, California  94720}
\end{center}

\vskip .5in

\begin{abstract}
A fit of amplitudes to the experimental branching ratios to two mesons
is used to construct a new estimate of neutral $D$ mixing which includes
$SU(3)$ breaking.  The result is dominated by the experimental uncertainties.
This suggests that the charm sector may not be as sensitive to new physics
as previously thought and that long-distance calculations may not be useful.
\end{abstract}


\vskip .3in
Submitted to {\it Physics Letters}\ {\bf B}

\end{titlepage}

\newpage
\renewcommand{\thepage}{\roman{page}}
\setcounter{page}{2}
\mbox{ }

\vskip 1in

\begin{center}
{\bf Disclaimer}
\end{center}

\vskip .2in

\begin{scriptsize}
\begin{quotation}
This document was prepared as an account of work sponsored by the United
States Government. While this document is believed to contain correct
information, neither the United States Government nor any agency
thereof, nor The Regents of the University of California, nor any of their
employees, makes any warranty, express or implied, or assumes any legal
liability or responsibility for the accuracy, completeness, or usefulness
of any information, apparatus, product, or process disclosed, or represents
that its use would not infringe privately owned rights.  Reference herein
to any specific commercial products process, or service by its trade name,
trademark, manufacturer, or otherwise, does not necessarily constitute or
imply its endorsement, recommendation, or favoring by the United States
Government or any agency thereof, or The Regents of the University of
California.  The views and opinions of authors expressed herein do not
necessarily state or reflect those of the United States Government or any
agency thereof, or The Regents of the University of California.
\end{quotation}
\end{scriptsize}

\vskip 2in

\begin{center}
\begin{small}
{\it Lawrence Berkeley Laboratory is an equal opportunity employer.}
\end{small}
\end{center}

\newpage
\renewcommand{\thepage}{\arabic{page}}
\setcounter{page}{1}

\section*{Introduction}

The prospects of probing new physics in $D^0$-$\bar{D}^0$ mixing has been
a topic recent discussion \cite{hewett} \cite{scipp}.
In this letter we will present a new estimate of $\Delta m_D$ due to
mixing via two-body hadronic intermediate states.
Because the couplings between the $D$ mesons and the possible intermediate
states are not all known, we will rely on the results of a fit to the
branching ratios \cite{dpaper}.
This fit utilizes a complete parameterization of the decays in the framework
of broken flavor $SU(3)$ and allows us to extract the (complex) couplings
needed to estimate $\Delta m_D$.
Intermediate states from the fit are limited to pseudoscalar-pseudoscalar (PP),
pseudoscalar-vector (PV), and vector-vector (VV).
We will find that the results are overwhelmed by their uncertainties.
These uncertainties owe their origin to the large (sometimes $\sim$30\%)
uncertainties on the experimentally measured branching fractions.
Other estimates, based on short-distance processes or on the heavy-quark
effective theory, give $\Delta m_D$ to be two orders of magnitude
smaller that our central values and uncertainties.
In the absence of cancellations among the hadronic modes,
our estimate allows the standard-model
contributions to $D^0$-$\bar{D}^0$ mixing to be close to the
current experimental limit $\sim 10^{-13}$ GeV \cite{pdg}.

\section{Formalism}

Weak interactions to second order in the coupling $G_F$ give an off-diagonal
part to the Hamiltonian which represents
$D^0 \leftrightarrow \bar{D}^0$ transitions.
Such a term is responsible for the mixing.
Its real part enters the mass matrix
and generates a mass difference between the two eigenstates.
In the absence of CP violation, we can write this as
  \begin{equation}
    {\cal H}_{\hbox{\scriptsize mass}} =
        \left( D^0 \; \bar{D}^0 \right)
        \left( \begin{array}{cc} M & \frac{1}{2} \Delta m \\
                                 \frac{1}{2} \Delta m & M
               \end{array} \right)
        \left( \begin{array}{c} D^0 \bar{D}^0 \end{array} \right).
  \label{lmass}
  \end{equation}
The mass eigenstates are
  \begin{equation}
  \begin{array}{l}
    D_1 = \frac{1}{\sqrt{2}} (D^0 + \bar{D}^0), \\
    D_2 = \frac{1}{\sqrt{2}} (D^0 - \bar{D}^0),
  \end{array}
  \label{eigen}
  \end{equation}
with masses
  \begin{equation}
  \begin{array}{l}
    m_1 = M + \frac{1}{2} \Delta m, \\
    m_2 = M - \frac{1}{2} \Delta m.
  \end{array}
  \end{equation}
The mass difference $\Delta m$ is a convenient parameter for the size of
the mixing.
It is the quantity whose measurement and calculation are important in
the question of probing new physics in $D^0$-$\bar{D}^0$ mixing.

Before continuing, we must draw the distinction between short- and
long-distance estimates to $\Delta m$.
Short-distance contributions come from the calculation of the ``box'' diagrams
in the quark model.
Long-distance contributions come from the dispersion due to intermediate
hadronic states.
In this case, the approach is only valid for internal momenta that are
below the scale at which QCD becomes nonperturbative.
Typically this scale is taken to be
  \begin{equation}
    \mu \; \sim \; 1 \; \hbox{GeV}.
  \end{equation}
The dependence on $\mu$ in our estimate will be explicit.

\section{Modelling the Couplings}

In order to construct our estimate, we need to know the couplings $D^0M_1M_2$
between the neutral $D$ mesons and the lowest-lying nonets of pseudoscalars
and vectors.
To this end, we have parameterized these couplings in the $SU(3)$ framework
with flavor-symmetry breaking by an octet.
The details can be found in \cite{dpaper}.
Here we will only sketch the calculation.

The particles are organized into $SU(3)$ multiplets.  The $D$ mesons ($D^0$,
$D^+$, $D_s$) form an antitriplet $\bar{\bf 3}$.
The pseudoscalars ($\pi^{\pm}$, $\pi^0$, $K^{\pm}$, $K^0$, $\bar{K}^0$,
$\eta$, $\eta'$) fit into an octet and a singlet.
The physical $\eta$ and $\eta'$ are mixtures of the octet and singlet pieces,
with mixing angle given by experiment \cite{cb}.
The vectors ($\rho^{\pm}$, $\rho^0$, $K^{*\pm}$, $K^{*0}$, $\bar{K}^{*0}$,
$\omega$, $\phi$) are likewise arranged, with a vector mixing angle that
is found in \cite{pdg}.

The charm-changing Hamiltonian is proportional to a product of quark currents
  \begin{equation}
    H_{\Delta C = 1} \sim \bar{u} \gamma^{\mu} (1-\gamma_5) q
                          \; \bar{q}' \gamma_{\mu} (1-\gamma_5) c,
  \label{hamiltonian}
  \end{equation}
where
  \begin{equation}
  \begin{array}{l}
    q = \cos \theta_C d - \sin \theta_C s, \\
    q' = \sin \theta_C d + \cos \theta_C s.
  \end{array}
  \label{quarks}
  \end{equation}
Since the quark-annihilation operators in $H$ transform as antitriplets, and
the $\bar{q}$ operators as triplets, we can expand $H$ in terms of $SU(3)$
representations.
The Hamiltonian is found to transform as {\bf 15} and $\bar{\bf 6}$.
The Clebsch-Gordan factors in this expansion and in the following were
calculated by computer \cite{cpc} \cite{isopaper}.

The amplitude for each decay of the type $D \rightarrow$ PP, PV, VV can
now be written as the sum of reduced matrix elements with appropriate
Clebsch factors.
For each of PP, PV, and VV there are 48 (complex) reduced matrix elements.
The number of parameters is too large to be fit by the available 45
measured modes and 13 modes with experimental limits \cite{pdg} \cite{purohit}.
Therefore, we must make some assumptions to limit their number.
first, we assume that corresponding reduced matrix elements of PP, PV, and
VV are proportional in magnitude.
This proportionality is represented by two new parameters
$A_{\hbox{\scriptsize PV/PP}}$ and $A_{\hbox{\scriptsize VV/PP}}$.
Second, we assume that the phase of each reduced matrix element is given
by the phase of the representation into which the daughter mesons are
contracted.
These are the $(\eta_1 \eta_1)_1$, $(\eta_1$P$)_8$, (PP)$_1$, (PP)$_8$,
(PP)$_{27}$,
$(\eta_1 \omega_1)_1$, $(\eta_1$V$)_8$, $(\omega_1$P$)_8$, (PV)$_1$, (PV)$_8$,
(PV)$_{8'}$, (PV)$_{10}$, (PV)$_{\bar{10}}$, (PV)$_{27}$,
$(\omega_1 \omega_1)_1$, $(\omega_1$V$)_8$, (VV)$_1$, (VV)$_8$, (VV)$_{27}$.
The phases then become new parameters.
With these assumptions, the number of linear combinations of reduced matrix
elements that contribute to any decay is reduced to 40.

There are now far fewer parameters than we began with.
The data constrain all of them, except for three combinations of reduced
matrix elements and the phases of $(\eta_1 \eta_1)_1$, $(\omega_1 \omega_1)_1$,
and $(\eta_1 \omega_1)_1$.
There are too many free parameters in the singlet-singlet cases, and so we
do not attempt to make any estimates of their values.
However, one of the remaining combinations of matrix elements
that involve the $D^0$ can be constrained by reasonable estimates of
one additional decay mode.
In order to see the effect of our lack of knowledge in this case,
we make two different estimates, called schemes A and B.
In the former we use
  \begin{equation}
    B(D^0 \arrow \eta K^0) =
        3 \tan^4 \theta_C \; B(D^0 \arrow \eta \bar{K}^0),
  \label{schemea}
  \end{equation}
and in the latter we take
  \begin{equation}
    B(D^0 \arrow \phi K^0) =
        3 \tan^4 \theta_C \; B(D^0 \arrow \phi \bar{K}^0).
  \label{schemeb}
  \end{equation}
The coefficient 3 is motivated by the size of the recently measured
mode $D^0 \rightarrow K^+ \pi^-$ \cite{cleodoublesupp}.
Interestingly, a coefficient of 1 does not result in a consistent fit.


\section{Estimate of $\Delta m$}

The long-distance contributions to $\Delta m$ arise from dispersive effects
involving intermediate states.
The two-body hadronic intermediate states were considered by \cite{donoghue}
are were estimated by
  \begin{equation}
  \begin{array}{lcll}
    \Delta m^{K^{\pm}, \pi^{\pm}}
        & = & \frac{1}{2 \pi} \ln \frac{m_D^2}{\mu^2}
            & \left[ \Gamma (D^0 \rightarrow K^+ K^-)
                   + \Gamma (D^0 \rightarrow \pi^+ \pi^-) \right. \\
        & & &\left.  - 2 \sqrt{ \Gamma (D^0 \rightarrow K^+ \pi^-)
                                 \Gamma (D^0 \rightarrow K^- \pi^+) } \right].
  \end{array}
  \label{eq:delmKpi}
  \end{equation}
Here $\mu$ is the cutoff discussed in Section 2.
Notice the implicit assumption that the couplings are relatively real.
If we insert the most recent values for these rates \cite{pdg}
\cite{cleodoublesupp}, we obtain
  \begin{equation}
    \Delta m^{K^{\pm}, \pi^{\pm}} = (\hbox{-}0.75 \hbox{ to } 0.29)
                                      \times 10^{-15} \hbox{GeV}.
  \label{eq:delmKpi:number}
  \end{equation}
The range of values is due to the uncertainty on the branching fraction to
$K^+ \pi^-$ \cite{cleodoublesupp}.
The purpose of this exercise is to show that large $SU(3)$ breaking may
give large long-distance contributions to neutral $D$ mixing.
Should the $SU(3)$ breaking be due only to the $K$-$\pi$ mass difference,
the rates would be related by
  \begin{equation}
         \begin{array}{lcl}
             \Gamma (D^0 \rightarrow K^+ K^-)
                            \Phi (K K)
                  & = & \Gamma (D^0 \rightarrow \pi^+ \pi^-)
                            \Phi (\pi \pi) \\
                  & = & \tan^2 \theta_C \Gamma (D^0 \rightarrow K^- \pi^+)
                            \Phi (K \pi) \\
                  & = & \cot^2 \theta_C \Gamma (D^0 \rightarrow K^+ \pi^-)
                            \Phi (K \pi),
         \end{array}
  \label{eq:gammaexact}
  \end{equation}
where $\Phi$ represents the phase-space corrections.
These corrections are on the order of a percent, and so in this case
  \begin{equation}
    \Delta m \simeq 10^{-17} \hbox{GeV},
  \label{eq:delmKpims}
  \end{equation}
a value consistent with other estimates, as discussed below.

We will make two improvements on this approach.
First, we will include all PP, PV, and VV intermediate states, with the
exception of the singlet-singlet states.
Second, we will allow the couplings to take complex values.
Equation \ref{eq:delmKpi} is replaced by
  \begin{equation}
    \Delta m_D^{\hbox{\scriptsize L-D}} =
        \frac{1}{2\pi} \ln \frac{m_D^2}{\mu^2}
        \times N \sum_I {\cal A} (D^0 \rightarrow I)
                        {\cal A}^* (D^0 \rightarrow \bar{I}),
  \label{eq:improved}
  \end{equation}
where $N$ is a normalization factor given by
  \begin{equation}
    \Gamma (D^0 \rightarrow K^- \pi^+)
        = N |{\cal A} (D^0 \rightarrow K^- \pi^+)|^2.
  \label{eq:normalization}
  \end{equation}
The coupling are extracted from the fit of the previous section.


Our estimates for $\Delta m$ are presented in the Table.
The estimates due to PV intermediate states are consistent
with zero and have uncertainties on the order of 50 $\times 10^{-15}$ GeV.
The estimates due to PP and VV intermediate states vary according to our
choice of estimate for the unconstrained modes (Equation \ref{schemea}
or \ref{schemeb}) and are at most one standard deviation from zero.
The source of the large uncertainties is the uncertainty with which we know
the individual branching fractions used to constrain our parameterization.
The contributions from individual modes enter with differing phases, but
the uncertainties are cumulative.
The result is that our attempt to estimate
$\Delta m$ is nearly overwhelmed by uncertainties.
It is also worth noting that these uncertainties are all greater than
the prior estimate which used only the modes $\pi^+ \pi^-$, $K^+ \pi^-$,
$K^- \pi^+$, and $K^+ K^-$ (see Equation \ref{eq:delmKpi:number}).

\section{Discussion}

We have found that a long-distance calculation of $\Delta m_D$ is dominated
by the experimental uncertainties.
The treatment of $D$-meson mixing by Donoghue {\it et al.}\ \cite{donoghue}
has underestimated these uncertainties.
On the other hand,
there also exist estimates based on the underlying quark processes.
The short-distance calculation based on the ``box'' diagrams has been done in
\cite{gaillard} \cite{burasbox} \cite{hagelin} \cite{donoghue} \cite{wolf}.
For example, \cite{donoghue} finds that
  \begin{equation}
    \Delta m^{\hbox{\scriptsize box}} = - \frac{G_F}{\sqrt{2}}
        \frac{\alpha}{4 \pi \sin^2 \theta_{\hbox{\scriptsize w}}}
        \cos^2 \theta_C \sin^2 \theta_C
            \frac{(m_s^2 - m_d^2)^2}{M_W^2 m_c^2}
            m_D F_D^2 (B_D - 2 B_D').
  \label{eq:deltambox}
  \end{equation}
Here $B$ and $B'$ are hadronic factors defined in \cite{donoghue}
and $F_D$ is the pseudoscalar decay constant:
  \begin{equation}
    \left< 0 | A_{\mu} | D^0 (p) \right> = i F_D p_{\mu}.
  \label{eq:fd}
  \end{equation}
If we assume that the hadronic factors $B \simeq B' \simeq 1$,
and take $F_D \simeq 300$ MeV \cite{pdg} and $m_s \simeq 250$ MeV,
then the contribution of the box diagrams to $\Delta m$ is
  \begin{equation}
    \Delta m^{\hbox{\scriptsize box}} \simeq 4.8 \times 10^{-17} \hbox{GeV}.
  \label{eq:deltambox:number}
  \end{equation}
In addition,
an estimate of the mass difference in the heavy-quark effective field theory
(HQEFT) was performed in \cite{georgiHQEFT} \cite{simmons}.
These arrive at estimates for the contributions of 4-, 6-, and 8-quark
operators:
  \begin{equation}
  \begin{array}{lcll}
    \Delta m_D^{\hbox{\scriptsize 4-quark}} & \simeq & 1
                                            & \times 10^{-17} \; \hbox{GeV}, \\
    \Delta m_D^{\hbox{\scriptsize 6-quark}} & \simeq & 2
                                            & \times 10^{-17} \; \hbox{GeV}, \\
    \Delta m_D^{\hbox{\scriptsize 8-quark}} & \simeq & 0.5
                                            & \times 10^{-17} \; \hbox{GeV}.
  \end{array}
  \label{eq:hqeftest}
  \end{equation}
These estimates replace both the long-distance calculation
and the box calculation.
In order to reconcile the HQEFT estimates with our approach, there must
be cancellation of the individual long-distance contributions to about
one percent.
In addition, it may be that the HQEFT estimate has underestimated the
$SU(3)$ breaking involved in the $D$ system.
The breaking of $SU(3)$ is known from the splittings of hadronic masses
to be at the level of 20-30\%.
However, in the charm system, we know from the branching fractions
that $SU(3)$ is broken at the 100\% level.
In the HQEFT estimate, only the quark-mass differences were used to break
the flavor symmetry.
We conjecture that this is the reason that the HQEFT estimate is on the same
order of magnitude as the box
and much smaller than the long-distance estimates.

The cancellation between contributions from different hadronic
intermediate states needed to reconcile the HQEFT approach with that
of the long-distance estimates would have to be among, rather than within,
the individual $SU(3)$ representations.
We can see this by considering the contribution due to the complete octet
of pseudoscalar mesons.
Although the singlet-singlet pieces of the PP decay modes are completely
unconstrained and the singlet-octet only partially constrained, the
octet-octet parts of the amplitudes are completely determined by data.
Therefore we are able to extract the octet-octet parts from the fit
without relying on the assumption of Equation \ref{schemea} or \ref{schemeb}.
It is then possible to
construct an estimate of $\Delta m$ that includes the
entire pseudoscalar octet, without the mixing in of singlet pieces.
We find that this contribution is
  \begin{equation}
    \Delta m_D^{\hbox{\scriptsize P octet}} = (9.6 \pm 2.2) \times 10^{-15}
        \; \hbox{GeV}.
  \label{eq:octet}
  \end{equation}
This estimate differs significantly from zero, and is therefore an indication
that cancellations among hadronic modes must be between the various
$SU(3)$ representations.

It is not possible for us to determine whether cancellation occurs within
the complete set of PP
intermediate states, including both $SU(3)$ octet and singlet pieces.
The expected size of the PP singlet-octet contribution can be seen in the
difference between Equation \ref{eq:octet} and the entries in the first
column of Table 1.
These singlet-octet contributions vary
by our estimation schemes for the unconstrained modes,
and are on the order of half of $\Delta m_D^{\hbox{\scriptsize P octet}}$.
We have no expectations on the size of the singlet-singlet contribution,
but can remark that it would be entirely due to $SU(3)$ breaking.
It is possible that PP modes will cancel among themselves, once
the singlet-singlet and singlet-octet pieces are fully included.
But should that not occur, then the only hope of cancellation would be
between the PP, PV, and VV modes taken together.

Due to the large uncertainties (sometimes $\sim 30\%$) on the experimentally
measured branching fractions, the uncertainties on our estimates of the
long-distance contribution are very large.
There are two things that could be learned from this.
First, the large uncertainties indicate that we are yet unable to make
a useful calculation of $\Delta m_D$ using intermediate hadronic states.
It is not reasonable for the experimental situation to improve to the point
where the uncertainties on $\Delta m_D$ reach the precision of the box or
HQEFT estimates in the near future.
Continued endeavors using one-particle (resonant) or three-particle
intermediate states are discouraged.
Second, in the absence of cancellations among the hadronic intermediate
states, the large range of possible values allows the standard-physics
contributions to $D$-meson mixing to be as large as
  \begin{equation}
    | \Delta m_D | = 10^{-13} \hbox{GeV}.
  \label{eq:largest}
  \end{equation}
This is near the experimental limit \cite{pdg}, and
therefore this process would be less likely to be useful as a probe of new
physics than previously thought \cite{hewett} \cite{scipp}.
We are left with the hope that a direct measurement of the $D$-meson
mass difference can be obtained and that these questions can be resolved.

\section*{Acknowledgements}

Many thanks go to Ian Hinchliffe for his valuable suggestions.

This work was supported by the Director, Office of Energy
Research, Office of High Energy and Nuclear Physics, Division of High
Energy Physics of the U.S. Department of Energy under Contract
DE-AC03-76SF00098.

\newpage
\mbox{ }
\vskip 1in

\tablecaption{}
        {Estimates of $\Delta m_D$.
         PP, PV, and VV refer to pseudoscalar-pseudoscalar,
         pseudoscalar-vector, and vector-vector intermediate states.
         Schemes A and B for estimating some doubly suppressed neutral modes
         are explained in Section 2.
         The first line neglects those modes.
         The last line includes the full pseudoscalar octet.
         In it, the octet-octet parts of $\eta\eta$, $\eta\eta'$, and
         $\eta'\eta'$ are included and the singlet-octet parts are excluded.
         All values are in 10$^{-15}$ GeV.}
\begin{center}
\begin{tabular}{c|ccc|}
  & PP & VV & PV \\
  \hline
  no estimates & 7.3 $\pm$ 3.9 & 19.1 $\pm$ 11.3 & -60.3 $\pm$ 63.3 \\
  scheme A     & 10.1 $\pm$ 4.4 & 25.5 $\pm$ 11.9 & -56.5 $\pm$ 63.9 \\
  scheme B     & 4.6 $\pm$ 4.5 & 14.7 $\pm$ 12.1 & -65.5 $\pm$ 63.9 \\
  $K^{\pm}$ and $\pi^{\pm}$ & 3.7 $\pm$ 1.3 & & \\
  full octet   & 9.6 $\pm$ 2.2 & & \\
  \hline
\end{tabular}
\end{center}

\end{document}